LSPM J1314: AN OVERSIZED MAGNETIC STAR WITH CONSTRAINTS ON THE RADIO EMISSION MECHANISM

James MacDonald & D. J. Mullan
Dept. Physics & Astronomy, University of Delaware, DE 19716, USA

Abstract

LSPM J1314+1320 (=NLTT 33370) is a binary star system consisting of two nearly identical pre-main sequence stars of spectral type M7. The system is remarkable among ultracool dwarfs for being the most luminous radio emitter over the widest frequency range. Masses and luminosities are at first sight consistent with the system being coeval at age ~80 Myr according to standard (non-magnetic) evolutionary models. However, these models predict an average effective temperature, 2950 ± 5 K, which is 180 K hotter than the empirical value. Thus, the empirical radii are oversized relative to the standard models by ≈13%. We demonstrate that magnetic stellar models can account quantitatively for the oversizing. As a check on our models, we note that the radio emission limits the surface magnetic field strengths: the limits depend on identifying the radio emission mechanism. We find that the field strengths required by our magnetic models are too strong to be consistent with gyrosynchrotron emission, but are consistent with electron cyclotron maser emission.

## 1. INTRODUCTION: OVERSIZED STARS

An increasing number of low mass, magnetically-active stars (those which are classified dMe, typically flare stars) have been found to have empirical radii which are oversized when compared to standard stellar models. The oversizing is statistically significant, reaching a $5\sigma$ effect in some cases. A principal goal of this paper is to discuss an evolutionary model which replicates the empirical oversizing in one particular star (LSPM J1314): our model is based on having magnetic fields of a certain strength present throughout the star. The main question of the paper will be: are the stellar fields, which our models require, consistent with the strong radio emission that is such a remarkable feature of LSPM J1314?

In the present section, we summarize the empirical results on oversizing of active stars that have emerged in recent decades as a result of a variety of observing methods. As far as we are aware, there are no reliable reports of significant *under*sizing in low-mass stars.

*1.1. Photometric evidence*

Mullan et al. (1989) reported on M dwarfs which were detected by the IRAS satellite: a sample of 55 highly active stars (with spectral types from dM0e to dM6e) were identified to have 12 micron (µm) fluxes which were systematically larger than the 12 µm fluxes for a sample of 31 inactive dM stars (between dM0 and dM7). For a given K magnitude, the dMe stars were found to be on average 1.7 times (i.e. 0.58 magnitudes) brighter at 12 µm than the dM stars at 12 µm. To the extent that the K magnitude is a measure of $T_{eff}$, the IRAS data suggest that active M dwarfs have larger radii than inactive M dwarfs by ~30% on average.

Hawley, Gizis, & Reid (1996) showed, for early-type M dwarfs (i.e., $V - I < 2.7$), that dMe stars are brighter on average than dM stars (in both the $V$ and $K$ bands) by $\Delta M$ = 0.48-0.66 mag. Moreover, in young clusters, the excess brightness of dMe stars over dM stars at the same color is even larger: $\Delta M$ is about 0.7 mag for Hyades stars and about 1.2 mag for stars in IC 2602. If color is a faithful proxy for effective temperature, $T_{eff}$, this indicates that the dMe stars are larger than the dM stars by 40 - 70%.

By combining $T_{eff}$ values, obtained by synthesis of infrared spectrophotometry, with bolometric luminosities, Leggett et al. (2000) determined stellar radii for a sample of 42 M dwarfs with uncertainties of 10% - 15%. When these data are plotted in a $T_{eff}$ *versus* radius diagram, we find that flare stars and variable stars tend to have larger radii than non-variable stars and have larger radii than the values that are predicted by stellar models (Mullan & MacDonald 2001).

Jackson et al. (2009) combined rotation periods with spectroscopic determinations of projected rotation velocity to determine mean radii of low-mass M-dwarfs in the young, open cluster NGC 2516. They found that the mean radii are larger than model predictions at a given absolute $I$ magnitude or $I - K$ color and also larger than measured radii of magnetically inactive M-dwarfs. The relative radius difference is correlated with magnitude, increasing from a few percent at $M_I = 7$ to greater than 50% for the lowest luminosity stars in their sample at $M_I \sim 9.5$.

*1.2. Eclipsing binary data*

Data from eclipsing binaries have greatly improved the precision with which empirical stellar radii can be determined. This leads to improved statistical confidence in the oversizing of a particular star. Here, we list results which typify how reliably the oversizing can currently be evaluated.

YY Geminorum (YY Gem), a member of the Castor Sextuplet, is a double line eclipsing binary, with $P_{orb} = 0.814$ d, containing two virtually identical M dwarfs. Torres & Ribas (2002) obtained for the mean mass and radius, the values $M = 0.5992 \pm 0.0047 M_\odot$, $R = 0.6191 \pm 0.0057 R_\odot$. By applying theoretical isochrones to the two A stars, Castor Aa and Castor Ba, Torres & Ribas (2002) determined the age of the Castor Sextuplet to be $370 \pm 40$ Myr. At this age, the mean component of YY Gem has a radius that is 5% – 15% greater than predicted by standard stellar evolution models. This is an oversizing of at least $5\sigma$.

CM Draconis (CM Dra) is an eclipsing binary, with $P_{orb} = 1.27$ d, containing two dM4.5 stars with masses of $0.23102 \pm 0.00089$ and $0.21409 \pm 0.00083$ $M_\odot$, and empirical radii of $0.2534 \pm 0.0019$ and $0.2398 \pm 0.0018$ $R_\odot$, respectively (Morales et al. 2009; Torres et al. 2010). The age of the system is constrained by the presence of common proper motion white dwarf companion to be $4.1 \pm 0.8$ Gyr (Morales et al. 2009), which places the stars on the main sequence. Comparing with stellar models having a range of ages and heavy element abundances, Morales et al. (2009) find that both components have radii that are larger than main sequence models predict by at least $0.01 R_\odot$. In view of the small statistical errors in the empirical radii, the 'bloating' of the radii of both components is, again, at least a $5\sigma$ effect.

LSPM J1112+7626 is an eclipsing binary system with component masses $M_1 = 0.395 \pm 0.002 M_\odot$ and $M_2 = 0.275 \pm 0.001 M_\odot$ in an eccentric ($e = 0.239 \pm 0.002$) orbit of period $41.03236 \pm 0.00002$ days (Irwin et al. 2011). Irwin et al. (2011) find that the sum of the component radii is oversized by $3.8^{+0.9}_{-0.5}\%$ compared to the theoretical model predictions, depending on the age and metallicity assumed. The long period of this system shows that radius oversizing is not confined to systems with very short orbital periods. A 65 day out-of-eclipse modulation is seen in I-band, and is probably due to rotational modulation of photospheric spots on one of the binary components. This spottedness is a clear signature of magnetic activity on stars that are oversized by at least 4$\sigma$.

HATS551−027 as an eclipsing binary with component masses and radii of $M_1 = 0.244^{+0.003}_{-0.003}$ $M_\odot$, $R_1 = 0.261^{+0.006}_{-0.009}$ $R_\odot$, $M_2 = 0.179^{+0.002}_{-0.001}$ $M_\odot$, $R_2 = 0.218^{+0.007}_{-0.011}$ $R_\odot$, and orbital period of ∼4.1 days (Zhou et al. 2015). HATS551−027 is one of few systems with both stellar components lying in the fully convective regime of very low mass stars. Zhou et al. (2015) find that the radius of HATS551−027A is consistent with models to 1$\sigma$, whilst HATS551−027B is oversized by 9 per cent at 2$\sigma$ significance. They measure the effective temperatures for the two stellar components to be $T_{eff,1}$ = 3190 ± 100 K and $T_{eff,2}$ = 2990 ± 110 K. These temperatures are lower than model predictions at the measured radius by 77 K and 142 K respectively.

LP 661-13 is a low mass binary system with an orbital period of $4.7043512^{+0.0000013}_{-0.0000010}$ days at a distance of 24.9 ± 1.3 parsecs (Dittmann et al. 2017). LP 661-13A has mass 0.30795 ± 0.00084 $M_\odot$ while LP 661-13B has mass 0.19400±0.00034 $M_\odot$. The component radii are 0.3226 ± 0.0033 $R_\odot$ and 0.2174 ± 0.0023 $R_\odot$, respectively. Dittmann et al. (2017) find that each component is slightly oversized compared to stellar models, and that this cannot be reconciled through age or metallicity effects.

*1.3. Brown dwarfs*

Radius oversizing is not restricted to lower main sequence stars but also has been reported in brown dwarf (BD) stars. Stassun et al. (2006) reported the discovery of the first eclipsing binary containing two brown dwarfs, 2MASS J05352184−0546085. In a follow-up paper, Stassun et al. (2007) found a "surprising reversal of temperatures" among the components. The primary BD component has a mass, $M_A$ = 0.0572 ± 0.0045$M_\odot$, some 60% larger than the mass of the secondary BD, $M_B$ = 0.0360 ± 0.0028$M_\odot$. Stassun et al. (2007) derived $T_{eff,A}$ = 2715 ± 100 K and $T_{eff,B}$ = 2820 ± 105 K, i.e. $T_{eff}$ for the primary is some 100 K *cooler* than $T_{eff}$ for the secondary. Even though the error bars of the $T_{eff}$ values overlap, the fact that the primary (with a mass that is some 60% larger than the secondary) is *not* clearly hotter than the secondary is certainly a "surprising" result. MacDonald & Mullan (2009) showed that the location of the secondary in the Hertzsprung-Russell diagram (HRD) is consistent with standard evolutionary models of BDs but the primary is too cool by at least 150 K compared to the model predictions. In other words, given the observed luminosity, the primary is too large by at least 10% compared to the models.

2. LSPM J1314: THE SUBJECT OF THIS PAPER

LSPM J1314+1320 (= NLTT 33370) is a binary star system containing two nearly identical pre-main sequence stars of spectral type M7. LSPM J1314+1320 (hereafter referred to as J1314) was first discovered to be a high proper motion star by Luyten (1979). The first clue to its binary nature was found by Law et al. (2006), who determined a binary separation of 130 mas. Lepine et al. (2009) measured the trigonometric parallax giving distance of 16.39 ± 0.75pc, and determined a spectral type M7.0e, with very strong Hα emission. The existence of strong Hα emission is a firm optical indication that the star is magnetically active.

However, in the case of J1314, it is not only in optical radiation that evidence for activity occurs. Radio emission is also prominent. McLean et al. (2011) discovered that J1314 is a source of persistent radio emission with a flat spectrum across a wide range of frequencies (1.43–22.5 GHz). The feature which draws attention to J1314 relative to other stars of similar spectral type is highlighted by McLean et al: it has "the most luminous radio emission over the widest frequency range detected from an ultracool dwarf (UCD) to date". Moreover, it is also one of the most X-ray luminous UCD's known (Williams et al. 2015).

McLean et al (2011) also discovered that the radio emission at 4.86 and 8.46 GHz is modulated with a period of ~3.9 hr, and the polarization at 4.86 GHz alternates between right-and left-handed circular polarization over a best-fit period of 3.8 ± 0.4 hr with an amplitude of 24 ± 10%. To explain this behavior, McLean et al. (2011) propose a large-scale dipolar magnetic field misaligned relative to the rotation axis and with opposite polarity at each pole. By combining the rotation period with $v \sin i$ measurements, they find the inferred radius is larger than models predict, by up to ∼30%.

Schlieder et al. (2014) confirmed the binary nature of J1314 by adaptive optics imaging, and obtained resolved near-infrared photometry and integrated light near-infrared spectroscopy. They also analyzed the integrated light optical spectra obtained by Lépine et al. (2009) and McLean et al. (2011). They estimated a system age of ∼30 – 200 Myr. From the infrared spectral energy distribution they determined $T_{eff}$ = 3200 ± 500 K and $T_{eff}$ = 3100 ± 500 K for J1314 A and B, respectively. The optical spectra showed weak, gravity-sensitive alkali lines and strong lithium 6708Å absorption, indicating a young age. From their analysis of the McLean et al. (2011) spectral data, Schlieder et al. (2014) determine the Li 6708Å absorption has equivalent width of ~460 mÅ.

Williams et al. (2015) have analyzed multi-epoch simultaneous radio, optical, Hα, UV, and X-ray observations of J1314, and note its extreme levels of magnetic activity, as it is the most radio-luminous, and one of the most X-ray luminous, ultra-cool dwarfs yet discovered. In the optical light curve, they find two periodicities, 3.7859 ± 0.0001 and 3.7130 ± 0.0002 hr, which they rule out as being due to differential rotation. Williams et al. suggest that the radio emission has 3 components: (i) short-lived flares where the polarization reaches 100%; (ii) bright emission in phase with optical emission; and (iii) another periodic component that appears in only one observing campaign. Williams et al. suggest that (iii) is a "gyrosynchrotron feature associated with large-scale magnetic fields and a cool, equatorial plasma torus." However, the occurrence of short-lived flares at all rotational phases suggest that small magnetic loops are also present. Williams et al. conclude that "the significant magnetism present in J1314 will affect its fundamental parameters": they suggest that the radii will be oversized by ∼+20%.

From analysis of their Keck adaptive optics astrometric monitoring and Very Long Baseline Array radio data from a companion paper (Forbrich et al. 2016), Dupuy et al. (2016, hereafter D16) determine component masses of $M_A$ = 0.0885 ± 0.0006 $M_\odot$ and $M_B$ = 0.0875 ± 0.0010 $M_\odot$, and a parallactic distance of 17.249 ± 0.013 pc. They find an orbital period of 9.58 years. D16 also find that the component luminosities (log $L_A/L_\odot$ = −2.616 ± 0.010, log $L_B/L_\odot$ = −2.631 ± 0.010) are consistent with the system being coeval at 80.8 ± 2.5 Myr, according to the evolutionary models of Baraffe et al. (2015). However, they find these the evolutionary models predict an average effective temperature, 2950 ± 5 K, that is 180 K hotter than the 2770 ± 100 K by a spectral type – $T_{eff}$ relation (Herczeg & Hillenbrand 2014) based on BT-Settl models (Allard et al. 2011, 2012). D16 suggest that the dominant source of the discrepancy is that the empirical radii are oversized by ≈ 13%. [Thus, as more precise data have become available, the reported oversizing has decreased from ~ 30% (McLean et al. 2011), to ~ 20% (Williams et al. 2015), to ≈ 13% in 2016.]

To summarize this section, the stars in LSPM J1314 have masses that are known to 1% or better, luminosities that are known to a few percent, and radii which are oversized by ≈ 13%. The goal of the present paper is to fit these data within the error bars using evolutionary models which include magnetic effects. If we can obtain an acceptable model, our solution will provide values of the vertical component of surface field strength $B_v$(surf). In an earlier paper, we checked on our theoretical values of $B_v$(surf) by comparing with X-ray data, which are known to be indirectly related to surface field strengths (MacDonald & Mullan 2014). However, in the case of LSPM 1314, there is no need to rely on indirect evaluation of fields: the observed radio emission may provide for the first time a direct consistency check on the surface fields in our magnetic model.

3. PROPOSALS FOR UNDERSTANDING OVERSIZED STARS

A number of models have been proposed to explain the oversizing of cool stars. In the context of the Leggett et al. (2000) sample and double line eclipsing binaries, Mullan & MacDonald (2001) proposed that the reddening of the dMe stars is due to the presence of their magnetic field. They considered a specific model based on the work of Gough & Tayler (1966) in which the magnetic field inhibits convective energy transport and leads to larger radii and lower effective temperatures than in non-magnetic stars. They noted that, in general, magnetically active stars should be larger than inactive stars. Chabrier et al. (2007) proposed a similar explanation to explain the temperature reversal in 2MASS J05352184−0546085 and, in addition, considered the impact of stellar dark spots.

In a recent paper (MacDonald & Mullan 2017, hereafter MM17), we discussed various effects of magnetic fields on stellar structure (including oversizing). The principal goal of MM17 was to examine if magnetic effects might lead to empirical discrepancies in the ages of various stars in a (presumably) co-eval group. Here, it is worthwhile to re-examine those reasons in the context of the system LSPM J1314, in order to determine if any of the reasons can be plausibly excluded from consideration in the present case.

(1) One possible source of an age inconsistency between different stars might be that a mass-dependent problem (of some kind) exists because of the way in which the ages of stars with different masses were determined. This needed to be considered in the case of MM17, where stars of widely different masses were involved. However, in the case of LSPM J1314, the masses of the two components are so similar (0.0885, 0.0875 $M_\odot$) that they may be regarded as identical, within the error bars. This suggests that reason (1) does not contribute significantly in the present case.

(2) A second possibility for the age inconsistency is that radius differences might occur for stars as young as a few times 10 Myr as a result of episodic accretion events (Baraffe et al. 2009; Baraffe & Chabrier 2010). Such differences might be interpreted erroneously as indicating ages that are too young. However, LSPM J1314 with an age of ~80 Myr is considerably older than the age (~10 Myr) of the system considered in MM17. For this reason, we consider that reason (2) does not contribute significantly to bloating of the stars in the present case.

(3) A third possibility is that low-mass stars might form at a different time than the high-mass stars. But this is irrelevant in the case of LSPM J1314, where the two stars in the binary have essentially identical masses.

(4) A fourth possibility, which was unintentionally omitted from MM17, is that overluminosity of a lower mass companion might in certain binaries with orbital periods of 5-10 days be due to tidal heating (Heller et al. 2010; Gómez Maqueo Chew et al. 2012). We recognize that it is possible for tidal heating to be significant in close binaries. However, since tidal forces scale as $a^{-3}$, the tidal effects are expected to be significantly weaker in wide binaries. In particular, in a binary as wide as LSPM J1314, with its 10 yr period (D16), tidal forces are expected to be weaker by factors of $P^{-2}$, i.e. by factors of order $10^5$ than in the case of orbital periods of 10 d. In view of this, we believe that reason (4) does not contribute significantly to bloating of the stellar radii in LSPM J1314.

(5) A fifth possibility, and the one we choose to examine here, is that magnetic fields alter the internal structure of the stars in LSPM J1314 sufficiently to lead to a $\approx 13\%$ increase in the stellar radius relative to the standard models.

4. MAGNETIC MODELLING OF LOW-MASS STARS

*4.1. Code updates*

Our code has previously been described in MacDonald & Mullan (2012, 2013, 2014). Here we note only some changes needed for the present application of the code. The options for treatment of the outer boundary conditions now include use of simple $T - \tau$ relations in separate atmosphere calculations. The general form of the relation is

$$\frac{4}{3}\left(\frac{T}{T_{eff}}\right)^4 = \tau + q(\tau), \tag{1}$$

where $q(\tau)$ is a generalized Hopf function. This relation is used to calculate the radiative gradient used in the mixing length theory of convection (Mosumgaard et al. 2016),

$$\nabla_{rad} = \left(\frac{d \ln T}{d \ln p}\right)_{rad} = \frac{3}{4ac} \frac{\kappa F_{tot} p}{g T^4}\left[q'(\tau) + 1\right], \tag{2}$$

where $F_{tot}$ is the total energy flux. In addition to the common Eddington approximation, for which $q(\tau) = 2/3$, options include the Krishna – Swamy (1966) relation

$$q(\tau) = 1.39 - 0.815 e^{-2.54\tau} - 0.025 e^{-30.0\tau}, \tag{3}$$

and, for limited ranges of $T_{eff}$ and $\log g$, the generalized Hopf functions determined by Trampedach et al. (2014) from 3D convection simulations.

*4.2. Inclusion of magnetic effects on convection*

Our approach to treating the inhibiting effects of a magnetic field on convection energy transport and mixing have been described earlier (MacDonald & Mullan 2009; Mullan & MacDonald 2010). The principal effect of the magnetic field is on the criterion for the onset of convection. In the fullest form of our model, we modify the standard Schwarzschild criterion to what we refer to as the Gough – Tayler – Chandrasekhar (GTC) criterion,

$$\nabla_{rad} > \nabla_{ad} + \Delta, \tag{4}$$

where, based on criteria derived by Gough & Tayler (1966) and Chandrasekhar (1961),

$$\Delta = \frac{\delta}{\theta_e} \min\left(1, \frac{2\pi^2 \gamma \kappa}{\alpha^2 \eta}\right). \quad (5)$$

Here, the magnetic inhibition parameter, $\delta$, is

$$\delta = \frac{B_v^2}{B_v^2 + 4\pi\gamma P_{gas}}, \quad (6)$$

where $B_v$ is the vertical component of the magnetic field, $\gamma$ is the first adiabatic exponent, and $P_{gas}$ is the gas pressure. Also $\theta_e = -\partial \ln \rho / \partial \ln T|_p$ is a thermal expansion coefficient, $\kappa$ is the thermal conductivity, $\alpha$ ($= l/H_p$) is the mixing length parameter, and $\eta$ is the magnetic diffusivity. As explained in MacDonald & Mullan (2009), the factor $\theta_e$ is introduced to account for the effects of deviations from an ideal gas on buoyancy and the factor proportional to $\kappa/\eta$ is included to account for the effects of finite electrical conductivity. To determine the convective energy flux, we replace $\nabla_{ad}$ by $\nabla_{ad} + \Delta$ in the mixing length theory. In addition, the magnetic pressure and magnetic energy density determined from the local value of $\delta$ are included in the stellar evolution equations.

The quantity $\delta$ is a local variable: in general, its numerical value may vary as a function of radial position in a star. Our specific radial profile for $\delta$ is that is constant down to a depth where it corresponds to a ceiling value of the vertical component of the magnetic field, which we denote by $B_{ceil}$. At deeper depth, $\delta$ is determined by equation (6) with $B_v$ set to $B_{ceil}$. Hence, our magnetic field profile is set by specifying two parameters: $\delta$ and $B_{ceil}$.

To apply our GTC criterion for instability, knowledge is needed of the magnetic diffusivity. For the contribution from collisions between electrons and charged particles, we use the electrical conductivity calculated by MacDonald (1991) for particles interacting through a static screened Coulomb potential. For the contribution from collisions between electrons and neutral hydrogen, we use the results of Temkin & Lamkin (1961) and Fon et al. (1978) for the total elastic scattering cross section. For collisions between electrons and $H_2$ molecules, we use the cross section data of Karwasz et al. (2003) tabulated by Yoon et al. (2008). Accurate fits to the resulting magnetic diffusivities over the temperature range $10^2$ K $< T <$ $10^5$ K are

$$\eta_{ea} = \frac{n_H}{n_e} \frac{933.695 \tau^{5/3}}{1 + 7.81515 \times 10^{-3} \tau + 1.18873 \times 10^{-3} \tau^2 - 2.83348 \times 10^{-4} \tau^3 + 1.74614 \times 10^{-5} \tau^4}, \quad (7)$$

where $\tau = T^{0.3}$, and

$$\eta_{em} = \frac{n_{H_2}}{n_e} \frac{10.289 \tau^{25/9}}{-0.03677 + 0.10092\tau - 0.03624\tau^2 + 0.00491\tau^3 - 2.4184 \times 10^{-4} \tau^4 + 3.14261 \times 10^{-6} \tau^5}, \quad (8)$$

where $\tau = T^{0.18}$. Here $n_{H_2}$, $n_H$ and $n_e$ are the number densities of hydrogen molecules, hydrogen atoms and electrons, respectively. At low temperatures, the free electrons are provided by the low first ionization potential elements Na, Ca and K, which are not included in our equation of state ionization balance calculation. We have explicitly added their contribution to $n_e$.

5. MAGNETIC MODELS OF LSPM J1314: NUMERICAL RESULTS

In this section, we explore how inclusion of magnetic inhibition of convection modifies the evolution of a model of mass appropriate for J1314, $M = 0.088$ M$_\odot$. For comparison purposes, we first consider a non-magnetic model that has outer boundary conditions determined from BT-Settl atmosphere models (Allard, Homeier & Freytag 2012; Allard et al. 2012; Rajpurohit et al. 2013). Since these atmosphere models use mixing length ratio, $\alpha = 1.0$, we adopt this value for the stellar interior. We also use the same composition as the atmosphere models, specifically that of Caffau et al. (2011). The evolutionary track for this model is shown by the black line in figure 1. We see that at the observed luminosity of J1314A, the predicted effective temperature is 2950 K. The spectral type–$T_{eff}$ relation of Herczeg & Hillenbrand (2014) gives $T_{eff} = 2770 \pm 100$ K.

Since the BT-Settl atmospheres do not allow for magnetic inhibition, we cannot, in a fully consistent way, include magnetic effects in our BT-Settl evolutionary models. Instead, we have looked for simple $T - \tau$ relations that give similar results to the BT-Settl models. We find that atmospheres constructed using a Krishna Swamy $T - \tau$ relation gives reasonable results for luminosities near that of J1314A. We then include magnetic effects consistently in the Krishna Swamy atmosphere and in the stellar interior.

The preferred explanation of Williams et al. (2015) for the periodic modulations of the optical light curve is the existence of magnetized cool spots, which requires strong coupling between magnetic field and the atmosphere. However, the temperatures in the outer parts of the star are sufficiently low that the degree of ionization is small. For example, in the outer $10^{-6}$ M$_\odot$ of the star, less than $10^{-5}$ of the electrons are free. For these reasons, we consider two cases; one in which the magnetic diffusivity is assumed zero, which gives the strongest coupling between field and matter, and another in which the magnetic diffusivity is derived based on the degree of ionization.

*5.1. Models in which the magnetic diffusivity is set equal to zero*

In this section, we consider models for the limiting case where the effects of finite electrical conductivity can be neglected altogether. That is, the electrical conductivity $\sigma$ is taken to be so high that the magnetic diffusivity, $\eta = c^2/4\pi\sigma$, is formally zero. The field remains rigidly frozen into the gas at all stages of the convective motions.

For this limiting case, Figure 1 shows the evolutionary paths (in red) in the HRD taken by models that use Krishna Swamy atmospheres, with each model having a fixed value of the magnetic inhibition parameter $\delta$ (see eq. (1.6)). In Figure 1, all red lines refer to models with $B_{ceil} = 10^4$ G. Also shown by a thick black line is the track for our non-magnetic model that uses BT-Settl atmospheres.

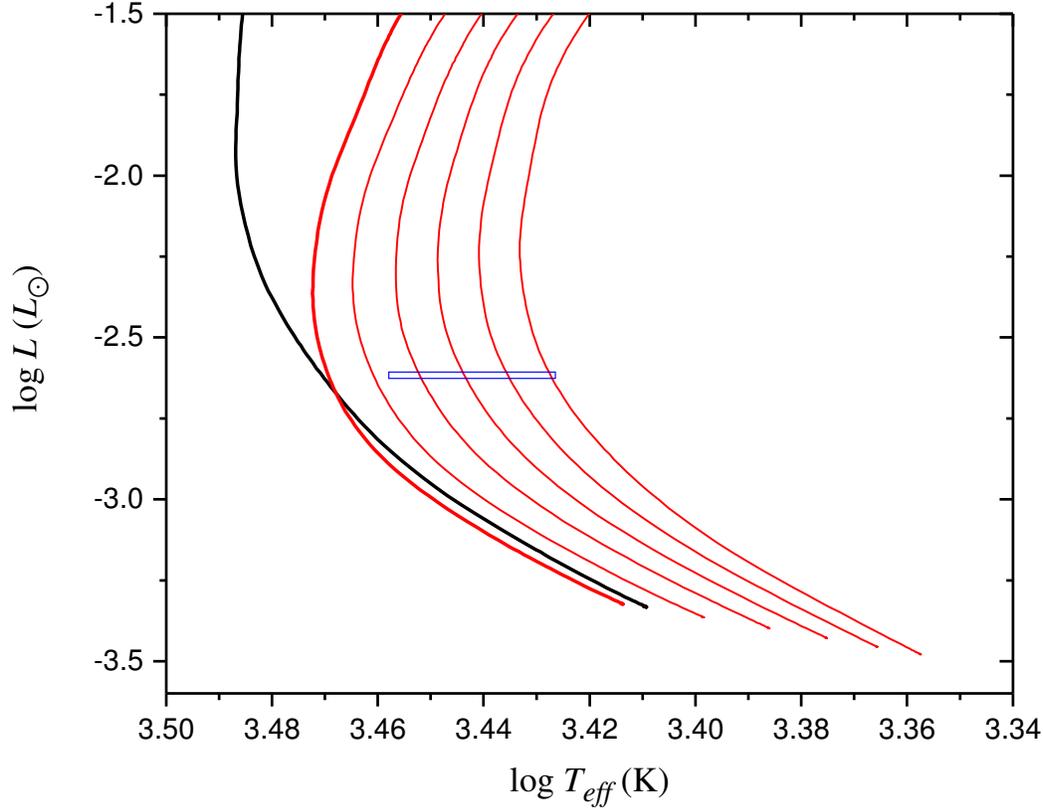

Figure 1. Evolutionary paths in the HRD of models for J1314A with $B_{ceil} = 10^4$ G. Each red curve refers to a magnetic model with a particular value of $\delta$. Values of $\delta$ vary from $\delta = 0.00$ on the left (thick red-line), to $\delta = 0.05$, in steps of 0.01 (thin red lines in order from left to right).

The blue rectangle in Fig. 1 spans the empirical $T_{eff}$ range obtained for J1314A by using the spectral type relation. The blue rectangle lies clearly to the right of the non-magnetic curves (thick black and thick red lines), i.e. the empirical $T_{eff}$ values are clearly *cooler* than the standard models. We see that the blue rectangle overlaps magnetic tracks which have $\delta$ values in the range $0.013 < \delta < 0.051$. Combining this range of values of $\delta$ with the gas pressure in the model at the photosphere, we find that the vertical component of the field strengths at the surface of the magnetic models inside the blue rectangle in Fig. 1 are $B_v$ (surf) = 440 – 880 G.

Figures 2 and 3 shows the evolutionary paths in the HRD taken by models for which we have selected $B_{ceil} = 10^5$ and $10^6$ G, respectively. For $B_{ceil} = 10^5$ G, the observational data are matched provided $0.009 < \delta < 0.032$, which corresponds to surface vertical field strengths of $B_v$ (surf) = 370 – 710 G. For $B_{ceil} = 10^6$ G, the observational data are matched provided $0.007 < \delta < 0.024$, which corresponds to surface vertical field strengths of $B_v$ (surf) = 330 – 610 G.

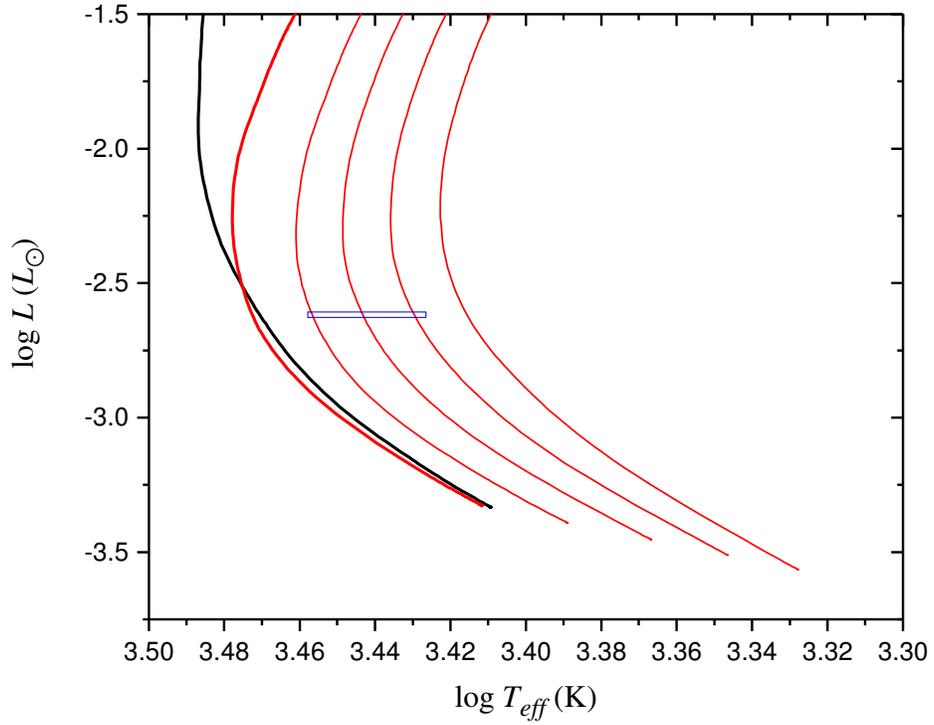

Figure 2. The evolutionary paths in the HRD taken by models of J1314A for which $B_{ceil} = 10^5$ G. Notation is the same as in Fig. 1, except that the range of $\delta$ is 0.00(0.01)0.04.

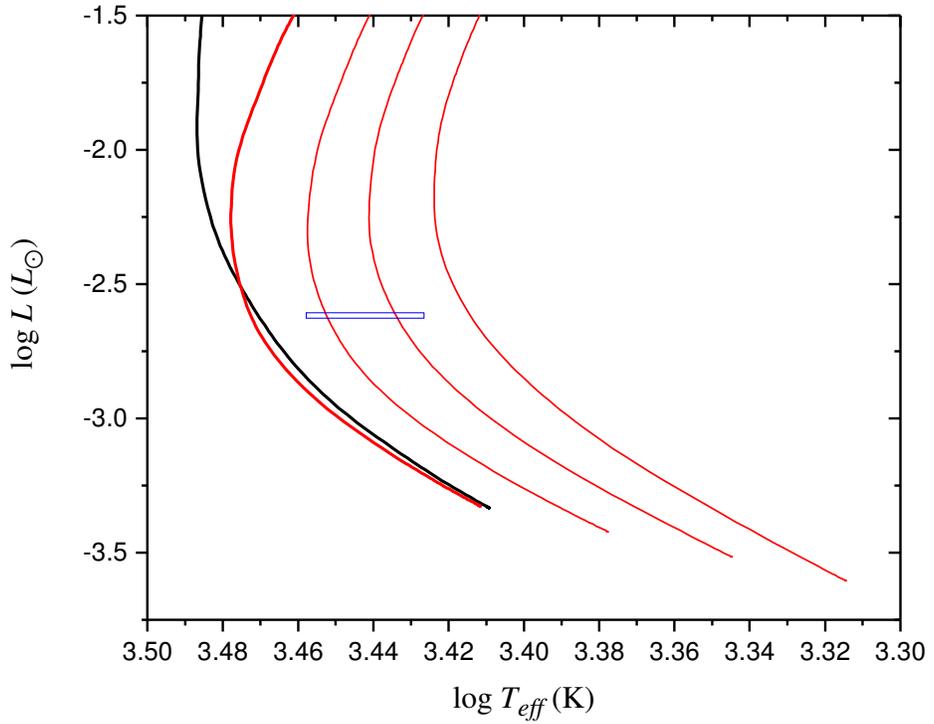

Figure 3. Evolutionary paths in the HRD taken by models of J1314A for which $B_{ceil} = 10^6$ G. Notation is the same as in Fig. 1, except that the range of $\delta$ is 0.00(0.01)0.03.

5.2. *Models in which the magnetic diffusivity takes on non-zero values*

In this section, we consider models for which the effects of finite electrical conductivity are included. In this case, the magnetic fields are able to slip out of the material to a certain extent during convective motions. The extent depends on the ratio between the linear dimensions of the convective cell and the linear distance $\approx \sqrt{\eta T}$ across which field lines can diffuse in the lifetime, $T$, of a convective cell. Because of this slippage of fields relative to the convective material, the fields must be stronger in order to replicate the same physical effects as were produced in the case of infinite conductivity. As a result, we expect that the solutions in this case will contain stronger fields than the solutions in Section 5.1.

Figure 4 shows the evolutionary paths in the HRD taken by GTC models for which $B_{ceil} = 10^4$ G.

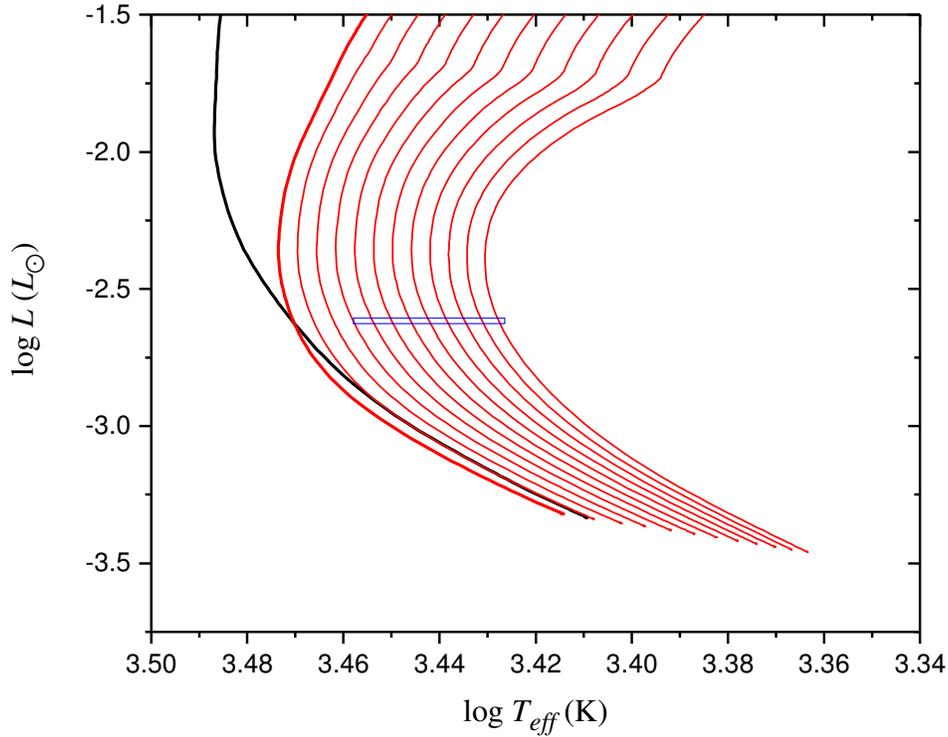

Figure 4. Evolutionary paths in the HRD taken by GTC models of J1314A for which $B_{ceil} = 10^4$ G. Notation is the same as in Fig. 1, except that the range of $\delta$ is 0.00(0.01)0.11.

The observational data are matched provided $0.03 < \delta < 0.11$, which corresponds to surface vertical field strengths of 630 – 1430 G. Because of the finite electrical conductivity, the required fields are (as expected) stronger than the fields that sufficed to replicate the 13% oversizing when the electrical conductivity is assumed infinite. In the present case, the fields are stronger by factors of about 1.5 compared to those obtained in the solutions presented in Section 5.1.

For our finite electrical conductivity models, we find that the required field strengths are only weakly dependent on $B_{ceil}$ (at least for $B_{ceil} < 10^6$ G). To see why, in figure 5, we plot $\Delta_1 = \delta/\theta_e$, and $\Delta_2 = \Delta_1 2\pi^2 \gamma \kappa / \alpha^2 \eta$ against mass for a model with $\delta = 0.03$ and $B_{ceil} = 10^4$ G.

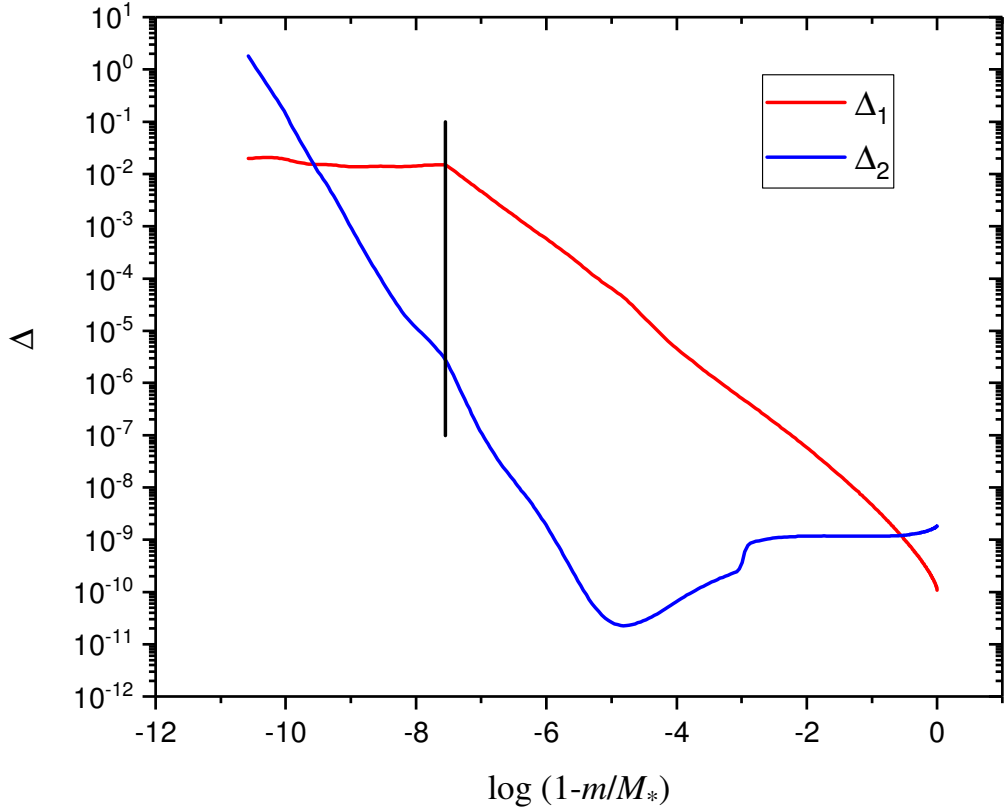

Figure 5. The parameter, $\Delta$, plotted against mass coordinate for a finite electrical conductivity model that matches the observational data for J1314A. The vertical line shows the depth at which the magnetic field ceiling of $10^4$ G is first reached.

Since $\Delta$ is defined (see equation 5 above) by $\Delta = \min(\Delta_1, \Delta_2)$, we see that the process of magnetic inhibition of convection (parametrized by the quantity $\delta$) dominates the magnetic solutions only for the outer $10^{-9}$ stellar masses or so. At greater depths, effects of magnetic diffusivity and thermal conductivity dominate the convective solutions, and the magnetic inhibition of convection is less important. In particular, the inhibition effects are negligible at depths where the ceiling is reached whether the ceiling value is $10^4$, $10^5$ G or greater.

*5.3. Model predictions of lithium depletion*

Since the two components of J1314 have similar masses and luminosities, it is reasonable to assume that they have the same degree of lithium depletion and the same equivalent width for the Li 6708Å absorption line. From the models of Baraffe et al. (2015), D16 determined that the fraction of lithium remaining in the primary and secondary components should be $0.12^{+0.05}_{-0.03}$ and $0.17 \pm 0.07$, respectively, which corresponds to a mean Li abundance of $A(Li) \approx 2.5$ [where $A(Li) = \log(N(Li)/N(H)) + 12$]. Using the relationship between Li I pseudo-equivalent widths and lithium abundance from the theoretical work of Palla et al. (2007), D16 determine that $A(Li) = 2.5$ corresponds to EW = 0.41–0.51 Å, in good agreement with the equivalent width measured by Schlieder et al. (2014). However, the Palla et al. (2007) calculations are for $T_{eff}$ values in the range 3100 – 3600 K and log g values of 4.0 and 4.5, whereas the components of J1314 have $T_{eff}$ ~ 2750 K and log g ~ 4.75. The Palla et al. (2007) results show that EW is increasing with decreasing $T_{eff}$ and increases with log g. Hence, D16 are likely underestimating the degree of Li depletion that is consistent with EW = 0.46 Å. Pavlenko et al. (1995) have calculated the relationship between Li equivalent widths and lithium abundance for models with $T_{eff}$ = 2500 K and 3000 K and log = 5. Combining the gravity dependence from the results of Palla et al. (2007) with the temperature dependence from Pavlenko et al. (1995), we estimate that EW = 0.46 Å corresponds to $A$(Li) = 2.1 and the fraction of Li remaining is 0.06, significantly lower than estimated by D16.

Figure 6 compares the degree of Li depletion as a function of $\delta$ for three sets of models that match the luminosity of J1314A. The black line corresponds to our GTC models (see section 5.2) and the red and green lines correspond to our models described in section 5.1 for $B_{ceil}$ = $10^4$ and $10^6$ G, respectively. The broken and solid horizontal lines shows the Li depletion determined by D16 and in this work, respectively.

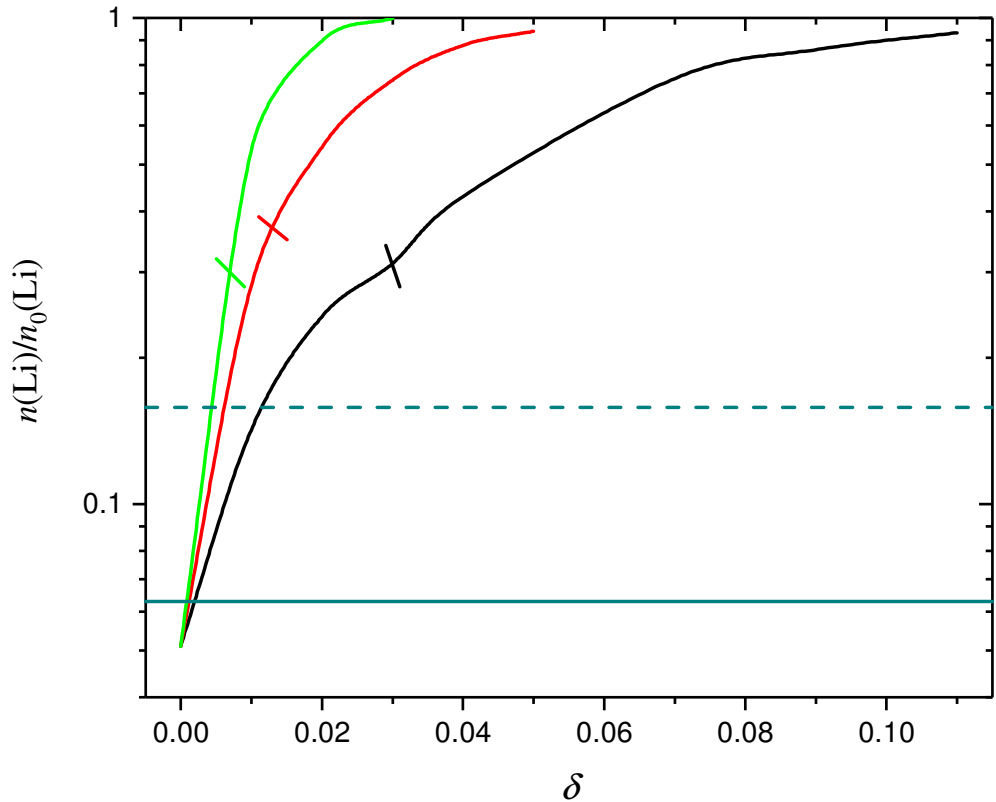

Figure 6. Lithium depletion for 3 models as a function of the magnetic inhibition parameter, $\delta$. The broken horizontal line shows the Li depletion determined from the equivalent width of the Li 6708Å absorption line by D16. The solid horizontal line shows our estimate of the Li depletion from the equivalent width. The short oblique lines mark the minimum value of $\delta$ for which the predicted and empirical $T_{eff}$ values are consistent,

From figure 6, we see that for models that are consistent with the observed luminosity, the predicted Li depletion is less than inferred from the Li 6708Å absorption line's equivalent width, independent of adopted values for $B_{ceil}$ and whether finite electrical conductivity effects are included or not. Because of its extremely high level of magnetic activity (Williams et al. 2015), we propose that the Li 6708Å absorption line is weakened due to photo-ionization of the Li ground state by chromospheric UV emission (Houdebine & Doyle 1997).

6. DISCUSSION AND CONCLUSIONS

In this paper, we have reported on magnetic models of stars where the masses and luminosities are equal to those of the components in J1314. By selecting a range of magnetic fields within the star, we have obtained magnetic models that achieve a satisfactory fit to the empirical radii, which are observed to be oversized by 13% relative to standard (non-magnetic) stellar models. We have therefore replicated the oversizing reported by Dupuy et al. (2016).

However, it was not merely to replicate the oversizing that we undertook our magnetic modeling. One of the outcomes of our models is that they provide a quantitative value for the strength of one component of the magnetic field (the vertical component) at the surface of the star. Our particular interest in paying attention to J1314 is that the physical interpretation of the radio emission from this star can be constrained by means of the magnetic field strengths predicted by our model. This is the first time (as far as we are aware) where a magnetic model of a star may be useful in setting constraints on the mechanism of radio emission.

We now compare two independent estimates of the magnetic field strengths on the surface of J1314. One is empirical, and is based on radio emission. One emerges from the theory of magnetic stars presented here.

(i) McLean et al. (2011) have reached the following conclusion. "If the radio emission is due to gyrosynchrotron emission, the inferred magnetic field strength is ~0.1 kG, while the electron cyclotron maser process requires a field of at least 8 kG". We note that the two different mechanisms of radio emission require the magnetic fields to differ in strength by almost a factor of 100. Independently, Williams et al. (2015) have reported on a "very bright Stokes V flare…suggesting a magnetic field strength of 2.1 kG" if the emission is an electron cyclotron maser. In these cases, the field strength refers to the total strength of the field, which will be larger than any individual component of the field (such as the one we calculate in our magnetic models).

(ii) In this paper, our magnetic solutions indicate that in order to fit the empirical 13% oversizing in radius, the vertical component $B_v$(surf) of the magnetic field at the surface of the star must have values that are in the following ranges.

(a) If there is no significant magnetic diffusivity, we find that $B_v$(surf) must have values of 440 - 880 G, 370 - 710 G, and 330 - 610 G if the ceiling field strength inside the star is $B_{ceil} = 10^4$, $10^5$, and $10^6$ G respectively. Note that, in order to replicate the empirical oversizing, the surface fields need to be stronger if we assume that the interior ("ceiling") fields are weaker. However, we stress that $B_v$(surf) is very insensitive to $B_{ceil}$: even when the latter increases by a factor of 100, $B_v$(surf) decreases by only 25% - 30%. It has been shown by Browning et al. (2016) that fields as strong as $10^6$ G (or stronger), can probably not survive as stable flux-ropes inside low-mass stars. In view of this, we regard our results for $B_{ceil} = 10^6$ G as setting extreme lower limits on $B_v$(surf). Therefore, in the absence of magnetic diffusivity, we consider that the vertical component of the surface field $B_v$(surf) has a lower limit of 0.37 - 0.88 kG. Since the total field strength must exceed this component, our non-diffusive models set a firm lower limit of 0.37 - 0.88 kG on the surface field strength on J1314.

(b) If magnetic diffusivity is operative, our models indicate find $B_v$(surf) = 630 - 1430 G for $B_{ceil}$ values greater than $10^4$ G, i.e. about 50% - 60% stronger than in the case of non-diffusivity. And once again, since our models yield only a value for the vertical component of the field, we conclude that our models require surface magnetic fields which are at least as strong as 0.63 - 1.43 kG.

Comparing our model results for $B_v$(surf) with the fields that have been reported for J1314 based on two distinct mechanisms of radio emission, it seems that our model results may help distinguish between the two possible mechanisms. Specifically, it is difficult to reconcile our results for the surface fields (> 0.37 - 1.43 kG) with the fields derived from gyrosynchrotron emission (~ 0.1 kG): our solutions have fields which are stronger by factors of at least 4, and possibly by more than a factor of 10. On the other hand, our magnetic solutions (which yield lower limits on surface field strengths) are entirely consistent with the field strengths that have been reported in association with electron cyclotron maser emission (2.1 kG, 8 kG).

As a result, our magnetic solutions for J1413 lead us to favor the electron cyclotron maser as the mechanism that explains the observed radio emission.

*Acknowledgments*
This work is supported in part by the NASA Delaware Space Grant.

References
Allard, F., Homeier, D., & Freytag, B. 2012a, Roy. Soc. London Philos. Trans. Ser. A, 370, 2765
Allard, F., Homeier, D., Freytag, B., & Sharp, C. M. 2012b, in EAS Pub. Ser. 57, eds. C. Reylé, C. Charbonnel, & M. Schultheis, 3
Baraffe, I., & Chabrier, G. 2010, A&A, 521, 44
Baraffe, I., Chabrier, G., & Gallardo, J. 2009, ApJL., 702, L27
Baraffe, I., Homeier, D., Allard, F., & Chabrier, G. 2015, A&A, 577, A42
Browning, M. K., Weber, M. A., Chabrier, G., & Massey, A. P. 2016, ApJ 818, 189
Caffau, E., Ludwig, H.-G., Steffen, M., Freytag, B., & Bonifacio, P. 2011, Solar Physics, 268, 255
Chabrier, G., Gallardo, J., & Baraffe, I. 2007, A&A, 472, L17
Chandrasekhar, S. 1961, in Hydrodynamic and Hydromagnetic Stability, International Series of Monographs on Physics, (Oxford: Clarendon)
Dittmann, J. A., Irwin, J. M., Charbonneau, D. et al. 2017, ApJ 836, 124
Fon, W. C., Burke, P. G., & Kingston, A. E. 1978, J. Phys. B: Atom. Molec. Phys., 11, 3
Forbrich, J., Dupuy, T. J., Reid, M. J., et al. 2016, ApJ, 827, 22
Gómez Maqueo Chew, Y., Stassun, K. G., Prša, A., et al. 2012, ApJ, 745, 58
Gough, D. O., & Tayler, R. J. 1966, MNRAS, 133, 85
Hawley, S. L., Gizis, J. E., & Reid, I. N. 1996, AJ, 112, 2799
Heller, R., Jackson, B., Barnes, R., Greenberg, R., & Homeier, D. 2010, A&A, 514, A22
Herczeg, G. J., & Hillenbrand, L. A. 2014, ApJ, 786, 97
Houdebine, E. R., & Doyle, J. G. 1995, A&A, 302, 861
Irwin, J. M., Quinn, S. N., Berta, Z. K., et al. 2011, ApJ, 742, 123

Karwasz, G. P., Brusa, R. S., and Zecca, A. 2003, in Photon and Electron Interactions with Atoms, Molecules and Ions, Landolt-Börnstein, New Series, Group I, Vol. 17, Pt. C, edited by Y. Itikawa (Springer, New York)
Krishna Swamy, K. S. 1966, ApJ, 145, 174
Law, N. M., Hodgkin, S. T., & Mackay, C. D. 2006, MNRAS, 368, 1917
Leggett, S. K., Allard, F., Dahn, C., Hauschildt, P. H., Kerr, T. H., & Rayner, J. 2000, ApJ, 535, 965
Lépine, S., Thorstensen, J. R., Shara, M. M., & Rich, R. M. 2009, AJ, 137, 4109
Luyten, W. J. 1979, New Luyten Catalogue of Stars with Proper Motions Larger than Two Tenths of an Arcsecond (NLTT), (CDS-ViZier catalog number I/98A; Minneapolis, MN: Univ. of Minnesota)
MacDonald, J. 1991, ApJS, 76, 369
MacDonald, J., & Mullan, D. J. 2009, ApJ, 700, 387
MacDonald, J., & Mullan, D. J. 2014, ApJ 787, 70
MacDonald, J., & Mullan, D. J. 2017, ApJ, 834, 67
McLean, M., Berger, E., Irwin, J., Forbrich, J., & Reiners, A. 2011, ApJ, 741, 27
Morales, J. C., et al. 2009, ApJ, 691, 1400
Mosumgaard, J. R., Aguirre, V. S., Weiss, A., Christensen-Dalsgaard, J., & Trampedach, R. 2016, arXiv:1610.07323v1
Mullan, D. J., & MacDonald, J. 2001, ApJ, 559, 353
Mullan, D. J., Stencel, R. E., & Backman, D. E. 1989, ApJ 343, 400
Palla, F., Randich, S., Pavlenko, Y. V., Flaccomio, E., & Pallavicini, R. 2007, ApJL, 659, L41
Pavlenko ,Y. V., Rebolo R., Mart´ın E. L., García López R. J., 1995, A&A, 303, 807
Rajpurohit, A. S., Reylé, C., Allard, F., et al. 2013, A&A, 556, A15
Temkin, A. & Lamkin, J. C. 1960 Phys. Rev., 121, 788
Torres, G., Andersen, J., & Gimenez, A. 2010, A&AR, 18, 67
Torres, G., & Ribas, I. 2002, ApJ, 567, 1140
Trampedach, R., Stein, R. F., Christensen-Dalsgaard, J., Nordlund, Å., & Asplund, M. 2014, MNRAS, 445, 4366
Williams, P. K. G., Berger, E., Irwin, J., Berta-Thompson, Z. K., & Charbonneau, D. 2015, ApJ, 799, 192
Yoon, J.-S., et al. 2008, J. Phys. Chem. Ref. Data, 37, 2
Zhou, G., et al. 2015, MNRAS, 451, 2263